\documentclass[prb,twocolumn,preprintnumbers,amsmath]{revtex4}
\usepackage{amsmath}
\usepackage{dcolumn}
\usepackage[dvips]{graphicx}
\usepackage{bbold}
\usepackage{booktabs}

\begin{document}
\bibliographystyle{apsrev}

\title{Reply to Comment on 
``Quantum phase transition in the four-spin exchange antiferromagnet"}

\author{Valeri N. Kotov}
\affiliation{Department of Physics, University of Vermont, 82 University Place, Burlington, VT 05405}
\author{D. X. Yao} 
\affiliation{Department of Physics, Boston University, 590 Commonwealth Avenue, Boston, MA 02215}
\author{A. H. Castro Neto}
\affiliation{Department of Physics, Boston University, 590 Commonwealth Avenue, Boston, MA 02215}
\author{D. K. Campbell}
\affiliation{Department of Physics, Boston University, 590 Commonwealth Avenue, Boston, MA 02215}

\begin{abstract}
We argue that the analysis of the J-Q model,  presented in Ref.~\onlinecite{kotov09}
 and based on a field-theory description  of coupled dimers,
  captures properly the strong quantum fluctuations tendencies, and the objections
 outlined in  Ref.~\onlinecite{isaev10} are misplaced.
\end{abstract}


\maketitle

In a recent paper, Isaev,  Ortiz, and Dukelsky  \cite{isaev10}
have questioned the interpretation of our results on the J-Q model
 \cite{kotov09}. In addition, they have argued that their hierarchical
 mean-field (HMF)  method  provides a more accurate description
of the phase diagram of the model \cite{isaev09}. In our view the
 conclusions of Refs.~\onlinecite{isaev10,isaev09} are highly questionable,
 and we address the two relevant issues below.

(1.) First, let us put our results (Ref.~\onlinecite{kotov09})
 in the correct perspective.  In the notation of  Ref.~\onlinecite{isaev10}, 
let $J_K$ be the 2-spin exchange
 and $-K$ be the 4-spin one; both $J_K$ and $K$ are assumed positive.
 These are the quantities we have used in our work, Ref.~\onlinecite{kotov09}
 (where we call $J_K$ simply $J$).
They are related to the  couplings $Q,J$ used in the original
 work of Sandvik \cite{sandvik07} via $Q/J = (K/J_K)/[1- K/(2J_K)]$.

 We prefer the $J_K,K$ units, because this way we can make the ratio of the  4- to the 2-spin
 term  arbitrarily large, whereas, for technical reasons, in the original work of  Sandvik
  this ratio has a maximum value of $(K/J_K) = 2$, which corresponds to
 $Q/J = \infty$. If $(K/J_K)$ is larger than 2, then this would mean $J<0$, but still
 the 2-spin exchange in those units, $(J + Q/2)$, is positive.
Now,  Sandvik's Monte Carlo (MC) result (confirmed also by later work \cite{melko,sandvik10}),
 gives the critical value $(Q/J)_c \approx 25$, meaning $(K/J_K)_c \approx 1.85$. 
 Our most advanced calculation finds the critical point at $(K/J_K)_c \approx 2.16$,
 which we argue to be an improvement  relative to our simple mean-field value, which is 
$(K/J_K)^{MF}_c \lesssim 1$.

Our main point is that while our  mean-field result  gives $(K/J_K)^{MF}_c \lesssim 1$ 
(i.e. $Q/J$ small, less than 2), our improvements, which take into account
the strong quantum fluctuations, lead to $(K/J_K)_c$ around 2,
 meaning that $|Q/J|\gg1$.  We call these "weak", and "strong" coupling regimes, respectively.
In this sense our value $(K/J_K)_c = 2.16$ is in "fairly good
 agreement with the MC", i.e. the critical point is firmly in the strong-coupling regime.
 
 The authors of Ref.~\onlinecite{isaev10} object to the fact that 
our $(K/J_K)_c = 2.16$ corresponds to $(Q/J)_c = -27$,
i.e. $J<0$,  which appears very far from the Monte Carlo: $ (Q/J)_c =+25$, i.e. $(K/J_K)_c = 1.85$. 
 However this a very misleading and incorrect  way to look at the results
(that's why we prefer the $K, J_K$ units).

 Since the transition is at large $|Q/J|$, it is irrelevant that $J<0$.
 Indeed,  the ratio of the 4- to the 2-spin term at our critical point, 
$(K/J_K)_c = 2.16$,  corresponds to the 4-spin term being $Q$, and the 2-spin term
 being $(-|J| +Q/2)$, with $Q/|J| =27$.
 Sandvik's result, $(K/J_K)_c = 1.85$, means that  the 4-spin term is $Q$, and the 2-spin term
 is $(J +Q/2)$, with $Q/J=25$.

 In both cases $|Q/J|\gg1$, so the sign of $J$ is irrelevant. This is what
 we call the strong-coupling regime, when $K/J_K$ is around 2, and
we describe the agreement between $(K/J_K)_c = 1.85$ and $(K/J_K)_c = 2.16$ as  "fairly good".  
 The two critical points  are fairly
 close, when things are put in the right context.  We have certainly  not achieved a perfect agreement
 with the Monte Carlo, but we have found the correct quantum fluctuations trend.

 Of course, as  pointed out in Ref.~\onlinecite{isaev10},
 the point $(K/J_K)_c = 2.16$ is outside the 
 range explored in the work of Sandvik \cite{sandvik07}, but the physics is expected
 to be the same, i.e. one simply penetrates  deeper into the  quantum disordered, gapped
 phase.

(2.) We also emphasize that our results were obtained under the assumption
 that the ground state is of the columnar dimer type, as argued in Ref.~\onlinecite{sandvik07}.
In our work we found that  this ground state is stable for $ K/J_K>(K/J_K)_c$,
 but we did not compare with other possible ground states,
 as it is quite difficult to compare ground state energies reliably \cite{kotov98},
 especially when the system exhibits strong fluctuations.
 
A severe problem of the HMF approach \cite{isaev09} is that 
 the critical point is at  $(Q/J)_c \approx 2$, i.e. $(K/J_K)_c \approx 1$, which is far off the Monte
 Carlo result. In fact the HMF critical point location is  close to the one
 we found in our ``simple" dimer mean-field framework \cite{kotov09},
 which we ruled out as unreliable. 
 Thus the hierarchical plaquette mean-field seems to place the critical point  firmly
 in the weak-coupling (small $Q/J$) region, in disagreement with  the Monte Carlo, and
 consequently it is highly unlikely that the HMF takes properly into account the
 strong quantum fluctuations present in the J-Q model.


\begin{thebibliography}{30}

\bibitem{isaev10} L. Isaev, G. Ortiz, and J. Dukelsky,  arXiv:1003.5205.

\bibitem{kotov09}
V. N. Kotov {\it et al.},  Phys. Rev. B {\bf 80}, 174403 (2009).

\bibitem{isaev09} L. Isaev, G. Ortiz, and J. Dukelsky,  J. Phys.: Cond. Matter {\bf 22}, 016006 (2010).


\bibitem{sandvik07}
A. W. Sandvik, Phys. Rev. Lett. {\bf 98}, 227202 (2007).

\bibitem{melko}
R. G. Melko and R. K. Kaul, Phys. Rev. Lett. {\bf 100}, 017203 (2008).

\bibitem{sandvik10}
A. W. Sandvik, Phys. Rev. Lett. {\bf 104}, 177201 (2010).


\bibitem{kotov98}
V. N. Kotov  {\it et al.}, Phys. Rev. Lett. {\bf 80}, 5790 (1998).



\end{thebibliography}
\end{document}